# Surface States Engineering of Metal/MoS$_2$ Contacts Using Sulfur Treatment for Reduced Contact Resistance and Variability

Shubhadeep Bhattacharjee*, Kolla Lakshmi Ganapathi*, Digbijoy N. Nath and Navakanta Bhat

Centre for Nano Science and Engineering, Indian Institute of Science, CV Raman Road, Bangalore 560012, India. E-mail: shubho@cense.iisc.ernet.in, navakant@ece.iisc.ernet.in

*Abstract*— **Variability and difficulty in achieving good Ohmic contacts are major bottlenecks towards the realization of high performance MoS$_2$-based devices. The role of surface states engineering through a simple Sulfur based technique is explored to enable reliable and superior contacts with high work function metals. Sulfur Treated (ST) multilayered MoS$_2$ FETs exhibit significant improvements in Ohmic nature, nearly complete alleviation in contact variability, ~2x gain in extracted field effect mobility, >6x and >10x drop in contact resistance and high drain currents with Ni and Pd contacts respectively. Raman and XPS measurements confirm lack of additional channel doping and structural changes, after Sulfur Treatment. From temperature dependent measurements, reduction of Schottky barrier height at Ni/MoS$_2$ and Pd/MoS$_2$ are estimated to be 81 meV and 135 meV respectively, indicating alteration of surface states at the metal/MoS$_2$ interface with Sulfur Treatment. Key interface parameters such as Fermi Pinning factor, Charge Neutrality Level and Density of Surface States are estimated using classical metal/semiconductor junction theory. This first report of surface states engineering in MoS$_2$ demonstrates the ability to create excellent contacts using high work function metals, without additional channel doping, and sheds light on a relatively unexplored area of metal/TMD interfaces.**

*These authors contributed equally. This work was supported in part by DeitY, Government of India under the Centre for Excellence in Nanoelectronics Phase II program.

S. Bhattacharjee, K. L. Ganapathi, D. N. Nath and N. Bhat are with the Centre for Nano Science and Engineering, Indian Institute of Science, CV Raman Road, Bangalore 560012, India (e-mail: shubho@cense.iisc.ernet.in; klganapathi@gmail.com; digbijoy@cense.iisc.ernet.in; navakant@ece.iisc.ernet.in).

## I. INTRODUCTION

The advent of 2D materials has opened alternate opportunities for transistor scaling, since traditional silicon technology has possibly hit technological and economic saturation beyond the sub-20 nm node [1-3]. Transition Metal Dichalcogenides (TMDs), unlike graphene, posses an intrinsic bandgap and hold promise for excellent electrostatic control in mitigating Short Channel Effects (SCE) [4,5]. In particular, $MoS_2$ (Molybdenum disulphide), with a bandgap of 1.2 eV in bulk, which is comparable to that of silicon, has gained prominence. FETs with $MoS_2$ channel material demonstrate high on to off current ratio and high mobility [6-8]. However, several key issues need to be addressed before the realization of TMDs-based FETs in viable technology. In this work we focus on engineering metal/$MoS_2$ interface to control variability and eliminate Schottky nature of contacts. Although preliminary reports suggested that the metal/$MoS_2$ interface forms Ohmic contacts with high work function metals [6,9,10], careful analysis confirmed the presence of a conspicuous Schottky barrier [11, 13]. Hence, the $MoS_2$ FET is essentially a Schottky Barrier Transistor and the intrinsic channel properties are heavily masked by contacts, thus rendering the scaling of the transistor quite inefficient [16].

Several interesting strategies have been formulated to circumvent this problem and can be categorized under: (1) Contact metal engineering (2) Doping of the channel to reduce the tunneling distance of carriers injected from metal to the channel. Under the first banner, several groups have tried a variety of contact metals including Sc[11], In[17], Al[17,18], Ti[11,13,17], Cr[7], Mo[19], Ni[11,20,21], Au[6,12,18], Pd[22] and Pt[11,18], with low work function metals forming relatively smaller Schottky barriers and enabling lower contact resistance. However, since the Fermi Level tends to get pinned just below the Conduction Band Minimum [23,24] at the interface, relying only on metal selection may not be a very effective approach. Besides the use of very low work function metals, such as Sc, is not compatible with existing CMOS technology. In the second strategy, doping of $MoS_2$ has been used to reduce the depletion width and aid tunneling current. Several demonstrations include the use of Potassium as adatom [25], polyethyleneimene (PEI) molecular doping [26], Plasma-assisted doping to form p-n junctions [27] and 12 hours of di-chloroethane dip which shows record current for Nickel contacted devices [20]. Apart from the degradation of doping with time, a key issue in most of these techniques is the lack of control, resulting in doping throughout the channel, instead of just below the contacts. A unique approach towards contact engineering was demonstrated by phase transformation of 1-H to 2-T metallic phase with 1 hour butyl lithium dip, yielding one of the lowest contact resistance values reported in literature [12]. Another major concern, although, not widely reported and discussed is that, the performance of $MoS_2$ FETs suffers from severe contact variability across the same wafer showing a large spectrum of contact nature from apparently Ohmic to completely Schottky [28,29]. This effect is typically pronounced in low Work Function (WF) metal contacts such as (Ti/Sc/Cr) which are used to achieve smaller Schottky Barrier Heights and reduced contact resistance. The gettering nature and reactivity with substrate of low WF metals even at low deposition pressures could be the primary cause for the same. Such large variability in contacts would result in low yield of devices which is unacceptable to technological and industrial demands. A recent report points at the role of surface defects and stoichiometric variations in accounting for device performance variability [24].

In this work we address a rather unexplored front in mitigating the issue of contacts to $MoS_2$; the possibility of surface preparation/treatment as a strategy to achieve predictable and superior performance contacts with high WF metals. Surface preparation has a long history in semiconductors and is reviewed in the reference [30]. We choose Sulfur Treatment owing to two primary considerations: first, the compatibility of Sulfur with the chemistry of $MoS_2$ and second, a historically large success rate in creating better contacts through Sulfur Passivation in several semiconductors like Germanium[31-33], Silicon[34,35] and other compound semiconductors[36]. In order to preserve the metal-semiconductor interface we select two high work function metals Nickel (5.0 eV) and Palladium (5.6eV) [24] which do not react with $MoS_2$ and are not prone to oxidation during deposition. We demonstrate that through an easy and inexpensive technique it is possible to alter and possibly uniformize the surface states at the metal/$MoS_2$ contacts.

## II. SULFUR TREATMENT & DEVICE FABRICATION

Back-gated FETs were fabricated by mechanical exfoliation of flakes from $MoS_2$ bulk crystal via the scotch tape method on 300 nm-$SiO_2$/p++ Si substrate (both acquired from Graphene Supermarket supplies). The $SiO_2$/Si wafer was subjected to Piranha clean and standard Acetone-IPA rinse prior to exfoliation. Next, the samples were segregated in two parts: Reference and Sulfur Treated (ST). For the samples marked as ST the entire flakes were subjected to Ammonium Sulphide solution $(NH_4)_2S$ [Sigma Aldrich Supplies, 40% solution in $H_2O$] treatment for 5 mins at a temperature of $50^0C$, followed by DI water rinse and $N_2$ blow-dry. The time and temperature parameters were optimized to ensure that no sulfur precipitation was observed. Acetone and IPA rinse were performed on both sets of samples to remove any residual and non-bonded chemical species from the surface and also remove traces of organic resist. Thickness layers of 5-7 nm were identified first using Optical Microscope and consequently confirmed by Atomic Force Microscopy (AFM, Bruker 500). Electron Beam lithography (Raith eLINE/Pioneer) was used to define patterns of fixed contact width and channel length of 1 μm each. Two sets of high work function metals Nickel (Ni) and Palladium (Pd), 60 nm each, were deposited which serve as contacts for both Reference and S-Treated samples using Techport e-beam evaporator at the pressure of $2x10^{-6}$ mbar and deposition rate of 2 Å/s. Aluminum (150 nm) was deposited as the back metal contact. All samples were annealed in vacuum ($2x10^{-6}$ mbar), inert Argon ambient ($1x10^{-3}$ mbar) at 400 °C for 1 hr. All electrical device parameters with the exception of temperature dependent studies are performed in ambient conditions with the Agilent B1500 Semiconductor Device Analyser. Temperature dependent measurements (100 K-400 K) were performed with the LakeShore Probe station.

## III. Results and Discussion

The primary observation in the output characteristics ($I_{ds}$-$V_{ds}$ (0-5V) with $V_{gs}$ sweep) was unambiguous: huge variability in the nature of contacts in case of Reference devices, results varying from apparently Ohmic to purely Schottky with different degrees of saturation [Figure 1(a)]. In contrast, the S-Treated devices demonstrated consistent Ohmic behavior with complete saturation [Figure 1(b)]. However, this contrast was clearly more prominent in Palladium devices because of a larger work function and Schottky Barrier Height (SBH) which is later explained quantitatively.

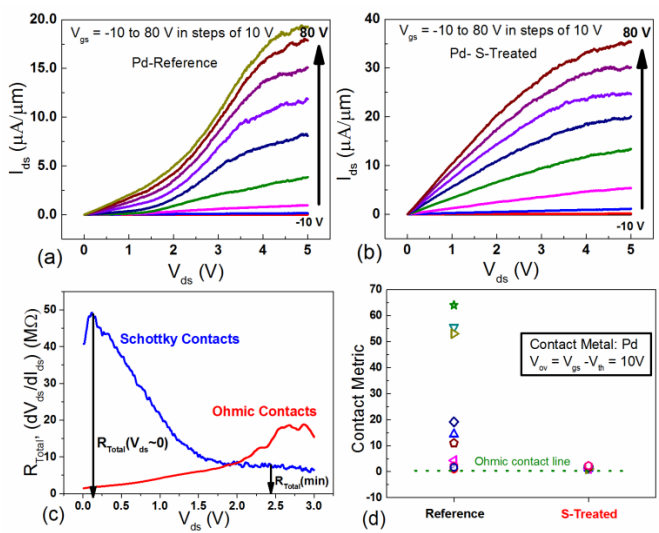

**Figure 1:** Contrast in nature of contacts in output characteristics for 5-7 nm MoS$_2$ flake: (a) Reference: Schottky Contacts with poor saturation. (b) ST device: Ohmic Contacts with excellent saturation (c) Contact Metric: derivative of output characteristics on sub threshold conditions: Ratio of R$_{Total}$ at V$_{ds}$ = 0 to minimum R$_{Total}$ (d) Huge variability in contact nature for Reference (20.71±24.49) against reliable Ohmic contacts in ST (1.43±0.45). Ohmic Contact Line at Contact Metric = 1, a guide to the eye.

To quantify the mitigation in variability and change of nature of contacts between Reference and ST samples a new 'Contact Metric (CM)' is defined (CM) = R$_{Total}$(V$_{ds}$ ~0)/R$_{Total}$(minimum), where R$_{Total}$ is small signal output resistance of the transistor, extracted from the derivative of I$_{ds}$-V$_{ds}$. The goal of the Contact Metric is to compare in a simple and intuitive fashion the 'Schottky nature' of contacts between Reference and ST devices. The appropriate region for extraction of R$_{Total}$ is in the 'ON' condition i.e. above threshold voltage where the linearity (Ohmic nature) of the output characteristics is essential. However, for large overdrive voltages even contacts with large barrier heights may give an impression of linearity due to excessive barrier width thinning, which can be misleading. Hence, the extraction of R$_{Total}$ is performed for a small overdrive voltage (V$_{ov}$ = V$_{gs}$-V$_{th}$). As evident in [Figure 1(c)], for purely Ohmic Contacts, the value for the Contact Metric is equal to 1. This is because the transistor is deeply in linear region for $V_{ds} \approx 0$, and then moves away from linear to saturation region with increasing $V_{ds}$, thus resulting in higher output resistance at larger $V_{ds}$ values. On the other hand, for Schottky contacts, the highest output resistance occurs for $V_{ds} \approx 0$, and then it decreases for larger $V_{ds}$, due to increasing electric field across the barrier. Hence, the contact metric value is substantially greater than 1. The Contact Metric was employed for 12 Palladium contacted Reference and ST devices each and the results are shown in [Figure 1(d)] measured for an equivalent small over drive voltage of $V_{ov} = 10V$. The mean value (20.71 for Reference and 1.43 for ST) is indicative of the deviation from purely Ohmic behavior (=1) and the standard deviation reflects the variability in different devices for the same wafer (24.49 for Reference vs 0.45 for ST). These numbers elucidate the large variability in the nature of contacts within the same wafer for $MoS_2$ FETs and its consequent suppression with Sulfur Treatment.

The improvement in contacts was also reflected on the Field Effect (FE) Mobility values (uncorrected for contact resistance, unlike in reference [20]) which demonstrated a ~2× improvement for both Nickel and Palladium [Figure 2(a)] contacted devices after ST. It is to be noted that the standard deviation in the mobility values does not attribute to the variability in contacts but flake to flake variation of $MoS_2$ layers in different devices, which affect both Reference and ST data equally [38]. For 1μm channel length FET, we obtain a saturation drain current value of 107 (167) μA/μm in ambient (vacuum), for a gate overdrive of 0.5V/nm [$(V_{gs}-V_{th})/t_{ox}$], which is on par with the best drain current value reported for Chlorine doped $MoS_2$ FETs [20], at twice the gate drive (1V/nm). These observations bring to the fore, the criticality of contacts in harnessing the intrinsic properties of $MoS_2$ based transistors.

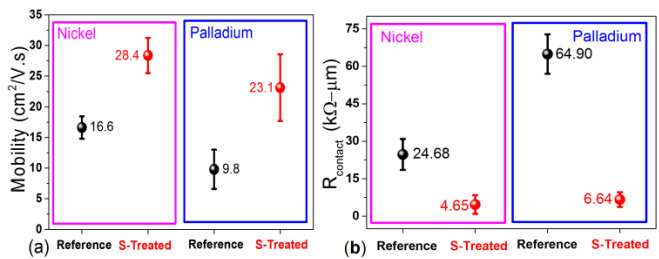

**Figure 2:** (a) Unmasking of FE Mobility in 5 best devices from each group with similar thickness: Nickel Contacte: Reference = (16.6±1.8) vs ST = (28.4±2.9) and Palladium Contacted: Reference = (9.8±3.2) vs ST = (23.1±5.4) (all in $cm^2V^{-1}s^{-1}$). (b) $R_{contact}$ calculated through the Y-Function Method demonstrates a ~6x (from 24.68±6.2 to 4.65±2.6) and ~10x (64.90± to 6.64±1.9) (all in kΩ-μm) reduction in Nickel and Palladium contacted devices respectively.

To gain further insight into the improved transistor characteristics with ST, Contact Resistance ($R_{contact}$) is extracted from two probe measurements using the Y-Function Method. The efficacy of the Y-Function Method in estimating $R_{contact}$ for multilayer $MoS_2$ FETs have been demonstrated [40, 41] by comparing the results to conventional 4-Probe and TLM measurements. Sulfur Treatment could significantly lower the $R_{contact}$ for Ni contacts by ~ 6x (from 24.68 kΩ-μm to 4.65 kΩ-μm) and Pd contacts by ~

10x (from 64.9 kΩ-μm to 6.64 kΩ-μm) [Figure 2(b)]. The values of $R_{contact}$ for ST devices are comparable to the lowest reported values of ~ 2 and 3 kΩ-μm with low WF metals such as Mo and Ti [19], where the channel below the contacts is unmodified. Furthermore, device-to-device variability in $R_{contact}$ was substantially lower for ST devices compared to Reference devices. With the Y-Function method it is also possible to calculate the 'intrinsic mobility' or the 'true' mobility of the channel not suppressed by contact resistance. The value of 'intrinsic' mobility for both ST and Reference devices lies in the range of 49-55 cm$^2$V$^{-1}$s$^{-1}$. This is in stark contrast to the 'extracted' FE mobility values which (includes the effect of contact resistance losses) show a ~2x improvement in case of ST devices, demonstrating the large influence of contacts on transistor performance. Furthermore, it illustrates that Sulfur Treatment does not cause any degradation in 'intrinsic' channel mobility.

## IV. Mechanism of Improved Contacts

To understand the drastic improvements in the contact performance of ST samples we look at two primary aspects in the metal/MoS$_2$interface: (1) Reduction in Schottky Barrier width and hence, tunneling distance as a result of doping and/or (2) Reduction in Schottky barrier height due to change in Density of Surface States.

To ascertain if the ST results in only surface level basal plane modifications or deep bulk level changes through doping, two strategies are adopted. First, Angle Resolved X-Ray Photoelectron spectroscopy (AR-XPS) measurements are performed on both Reference and ST samples at 0, 30 and 60 degrees before metallization. Several reports have suggested a 0.6-1.2 eV shift in Binding Energy, for both Mo and S peaks, positive shift for n-type doping and negative shift for p-type doping [20, 25, 27].The results for all angles show no alteration in Binding Energy or nature of Mo 3d/5d and S2p peaks, providing unambiguous evidence that MoS$_2$ does not undergo bulk level changes in electronic configuration after ST [Figure 3(a)]. Furthermore, micro-Raman [Figure 3(b)] and Photolumiscence performed on same flakes before and after ST demonstrated absence of chemical/structural change post ST. These results could be energetically reconciled considering that the relatively low temperature (50 $^0$C) and time (5 mins) of Sulfur Treatment energetically favor diffusion of Sulfur Vacancies [37] on the top basal plane rather than diffusion of S atoms through the Van der Waals layers and/or alterations to a stable MoS$_2$ bond (ΔG$^0$= -225.9 kJ/mol). It has to be categorically stated that these techniques can provide insight regarding changes to MoS$_2$ bulk film only and not the top basal plane.

The Schottky nature of metal/ MoS$_2$ contacts has been widely reported in literature [11, 14, 41]. [Figure 4(a)] demonstrates the change in current conduction from Thermionic Emission to Tunneling with increase in gate voltage as measured in the temperature dependent transfer characteristics [15].

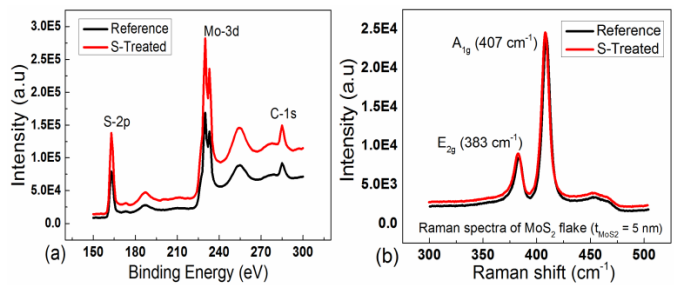

**Figure 3:** (a) X-Ray Photoelectron Spectroscopy (b) micro-Raman analysis for Reference and ST, no shift in Mo-3d/S-2p peaks, hence no doping or structural alteration as a result of ST.

The conventional Thermionic emission model ($I_{ds}=AT^2\exp(-\Phi_b/k_BT)[(\exp(qV_{ds}/k_BT)-1])$) was used for extraction of barrier height for different gate voltages[11, 14]. Where, A is the Richardson's Constant, T is the temperature in Kelvin, $\emptyset_B$ is the Schottky Barrier Height (SBH), $k_B$ is the Boltzmann's constant and $V_{ds}$ is the source to drain voltage fixed at 100 mV. For gate voltages deep in the sub-threshold region, the drain current is dominated by thermionic component which results in a linear dependence of $\emptyset_B$ with $V_{gs}$. At the onset of Flatband condition, the tunneling current contribution begins to play a significant role and non-linearity is observed in $\emptyset_B$ vs $V_{gs}$ plot [Figure 4(b)]. The Barrier Height extracted at this flat band condition is termed the 'true' SBH [11]. The true SBH is measured for Nickel and Palladium contacted devices as shown in [Figure 4(c)] for Reference and ST devices. For Nickel contacted devices, barrier height was measured to be 260.30±24.35 meV and 179.25±18.88 meV for Reference and ST devices respectively, recording a difference of nearly 81 meV. Palladium, as expected demonstrated a larger reference SBH of 331.82±17.86 meV which dropped to 196.08±8.64 meV after ST, with a difference of nearly 135 meV. Furthermore, reduced Schottky Barrier Height resulting in better charge injection into the channel was noted by early turn on of ST devices and a ~20 V left shift in Flatband Voltage ($V_{FB}$) as also demonstrated by S. Das et.al with different metals [11].

Temperature dependent mobility measurements present further benefits of Contact improvements. The experimentally extracted values with and without ST are compared against a model considering two dominant scattering mechanisms, Remote Impurity Scattering and Optical Phonon Scattering [Figure 4(d)] replicating the models used by S. Kim et.al [13]. Substantially reduced $R_{contact}$ losses in Sulfur Treated devices are evident by a lower suppression of channel mobility and the extracted mobility approaching the theoretical model. Furthermore, ST devices exhibit lower mobility degradation vs. temperature ($\propto T^{-1.01}$) compared to Reference devices ($\propto T^{-1.21}$) extracted for T ≥ 250K.

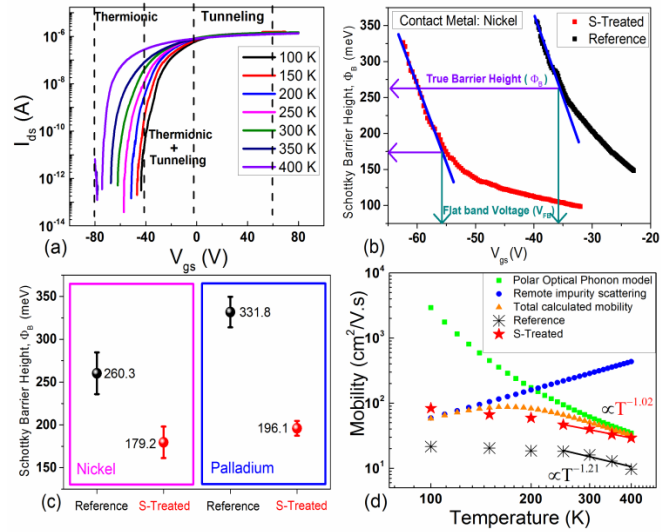

**Figure 4:** (a) Temperature dependent transfer characteristics demonstrating shift from thermionic regime to tunneling regime with increase in $V_{gs}$. (b) SBH vs $V_{gs}$ deviation from linear behaviour marks the onset of Flat Band Condition where the 'true' SBH is measured for Ni ST and Ni Reference (c) Nickel contacted devices: SBH recorded to be 260.3±24.3 meV for Reference and 179.3±189 for ST; Palladium contacted devices: larger reference barrier height of 331.8±17.8 meV which dropped to 196.1±8.6 meV after ST. (d) Temperature dependent mobility values are compared against a model considering two dominant scattering mechanisms, Remote Impurity Scattering and Optical Phonon Scattering

## V. Estimation of Surface parameters with and without Sulfur Treatment

The fundamental charge neutrality equation was used to determine the change in surface states before and after S-Treatment [42]:

$$\emptyset_{Bn} = S(\emptyset_m - \chi) + (1-S)(E_g - \emptyset_0) \qquad (1)$$

$$D_{it} = \frac{(1-S)\varepsilon_i}{S\delta q^2} \qquad (2)$$

Where, the known parameters in the equation (1) are, $q$ the elementary charge, $\emptyset_{Bn}$ the extracted SBH, $\emptyset_m$ the Work Function of the contact metal, $\chi$ and $E_g$ the Electron Affinity and bandgap of MoS$_2$ taken to be 4.1 eV and 1.2 eV respectively. The two unknown parameters, the Pinning Factor, $S$ which can take values from 0 (complete pinning) to 1 (no pinning)and the Charge Neutrality Level above the Valence Band, $\emptyset_0$ are calculated by simultaneously solving the equation (1) before and after ST, for Nickel and Palladium devices with work function of 5.0 and 5.6 eV respectively. The Pinning Factor is used to determine the Density of Surface States ($D_{it}$) by using equation (2), assuming the interface permittivity ($\varepsilon_i$) to be equal to vacuum and the interfacial layer width ($\delta$) to be in the order of 0.4-0.5 nm. The results are summarized in the Table 1 below:

| Parameters | Fermi Pinning Factor (S) | Charge Neutrality level ($\emptyset_0$)(in eV) | Density of Surface states ($D_{it}$)(in cm$^{-2}$eV$^{-1}$) |
|---|---|---|---|
| Reference | 0.122 | 1.029 eV | 7.98×10$^{13}$ |
| S-Treated | 0.028 | 1.041 eV | 3.82×10$^{14}$ |

**Table 1:** Key surface parameters (Fermi Pinning Factor, Charge Neutrality Level, Density of Surface states) as extracted on Reference and S-Treated devices.

Several key aspects of metal/MoS$_2$ contacts could be encapsulated through these results [Figure 5]. First, we confirm and quantify strong Fermi Level Pinning in MoS$_2$ contacts, with a pinning factor ($S$) of 0.122, nearly identical to the value (0.1) measured by S. Das et. al [11] with 4 metals. Furthermore, if we try to evaluate the expected SBH of Ti contacts using the Fermi-pinning Equation (1): $\emptyset_B(Ni) - \emptyset_{Bn}(Ti) = S\emptyset_m(Ni) - \emptyset_m(Ti)$ ,$\emptyset_m(Ni) = 5.0\ eV$ ,$\emptyset_m(Ti) = 4.3\ eV\emptyset_{Bn}(Ni) = 260\ meV$ we get $\emptyset_{Bn}(Ti) = 174.6$ meV which is nearly equal to $\emptyset_{Bn}$ of Ni(ST). This provides clear indication that ST is an effective method to mimic contacts with low work function metals and still achieve very low contact variability. Second, the proximity of the Charge Neutrality Level ($\emptyset_0$) to the Conduction Band Minimum matches theoretical predictions [23, 24].Third, while in traditional semiconductors, ST is known to "passivate the surface" reducing surface states and depinning the Fermi level, metal/MoS$_2$ contacts demonstrate the contrary.

A plausible model to explain these results could be the presence of spatially non uniform sub-oxides of Molybdenum (MoO$_x$) on the basal plane. The oxides when sulfurised or etched away by the proposed treatment produce uniformly pristine MoS$_2$ surface on which contacts are formed. It is well known that sub-oxides in traditional semiconductors aid in the de-pinning the Fermi level. Hence, the removal of these sub-oxides leads to strongly and reliably pinned n-type contacts governed entirely by surface states. However, this model remains to be accurately tested with the help of advanced atomic level imaging techniques such as Scanning Tunneling Microscopy/Spectroscopy.

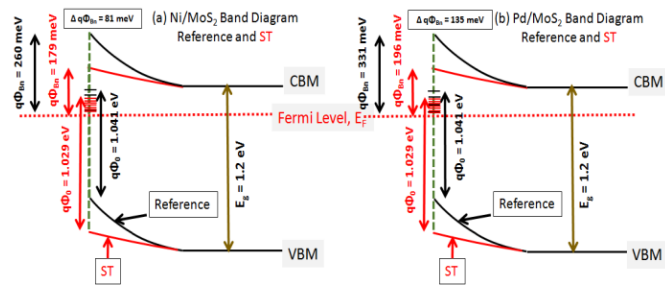

**Figure 5:** Band Diagrams elucidating the impact of S-Treatment of surface states on (a) Ni/MoS$_2$ and (b) Pd/MoS$_2$ contacts.

## VI. Conclusions

In summary, a novel method of surface engineering, in the context of $MoS_2$ back gated FETs, is demonstrated through Ammonium Sulphide treatment in a bid to systematically improve contact performance and reliability with high WF metals [Table 2]. ST devices show consistent Ohmic characteristics with good saturation against a range of Schottky to Ohmic behavior with poor saturation in case of Reference samples. Controlling contact variability is essential for improving device yield for CMOS technology. For the four sets of devices; Ni(ST), Pd(ST), Ni(Ref), Pd(Ref), while extracted field effect mobility values demonstrate inverse correlation with measured barrier height, contact resistance demonstrates direct correlation, signifying the importance of contacts in device performance. It is established through material characterization techniques that improvement could not be attributed to bulk doping effects. Temperature dependent transfer characteristics measurements provided clear evidence of reduction in Schottky barrier height and enhanced charge injection into the channel with ST. Key interface parameters with and without ST are determined using classical M/S theory elucidating that contrary to traditional semiconductors, following ST, metal/$MoS_2$ contacts are governed entirely by surface states. By uniformly controlling these surface states; it is possible to engineer high performance reliable Ohmic contacts relatively insensitive to differences in metal work functions. This is evident for 2 metals Ni and Pd where the difference in barrier heights reduces from 71 meV to 17 meV and contact resistance difference drops from 40 k$\Omega$-$\mu$m to 2 k$\Omega$- $\mu$m after ST. This study also reveals that it would be difficult to harness the true potential of transistors on TMD materials, without paying close attention to metal/TMD interfaces.

### Acknowledgment

The authors would like to thank DeitY, Govt. of India, for funding support through the project Centre for Excellence in Nanoelectronics – Phase II. This publication is an outcome of the R&D work undertaken in the project under PhD scheme of Media Lab Asia. The authors would like to acknowledge the National Nano Fabrication Centre (NNFC) and the Micro and Nano Characterization Facility (MNCF) at the Centre for Nano Science and Engineering for access to fabrication and characterization facilities.

| Performance Metrics | Contact Variability Metric (*Ohmic Contact =1*) | $I_{ds}$ (saturation) $L_{ch}$ =1$\mu$m, (in $\mu$A/$\mu$m) | Extracted Field Effect Mobility (in $cm^2V^{-1}s^{-1}$) | | $R_{contact}$ (in k$\Omega$-$\mu$m)) | | Schottky Barrier Height (in meV) | | Pinning factor | Mobility $\propto T^{-\gamma}$ $\gamma=$ |
|---|---|---|---|---|---|---|---|---|---|---|
| | | | *Ni* | *Pd* | *Ni* | *Pd* | *Ni* | *Pd* | | |
| Reference | 20.71 | 90-100 | 16.6 | 9.8 | 24.68 | 64.90 | 260.3 | 331.8 | 0.122 | 1.21 |
| ST | 1.43 | 160-170 | 28.4 | 23.1 | 4.65 | 6.64 | 179.2 | 196.1 | 0.028 | 1.02 |

Table 2: A summary of performance metrics extracted for Reference and Sulfur Treated FETs